
\input phyzzx

\def\pr#1#2#3#4{Phys. Rev. D{\bf #1}, #2 (19#3#4)}
\def\prl#1#2#3#4{Phys. Rev. Lett. {\bf #1}, #2 (19#3#4)}

\def\np#1#2#3#4{Nucl. Phys. {\bf B#1}, #2 (19#3#4)}
\def\pl#1#2#3#4{Phys. Lett. {\bf #1B}, #2 (19#3#4)}

\nopagenumbers
\line{\hfil CU-TP-577}
\vglue .5in
\centerline {\twelvebf  Vacuum Decay in Theories with Symmetry
Breaking by Radiative Corrections}
\vskip .3in
\centerline{\it Erick J. Weinberg}
\vskip .1in
\centerline{Physics Department, Columbia University}
\centerline{New York, New York 10027}
\vskip .4in
\baselineskip=14pt
\overfullrule=0pt
\centerline {\bf Abstract}

     The standard bounce formalism for calculating the decay rate of a
metastable vacuum cannot be applied to theories in which the symmetry
breaking is due to radiative corrections, because in such theories the
tree-level action has no bounce solutions.  In this paper I derive a
modified formalism to deal with such cases.  As in the usual case, the
bubble nucleation rate may be written in the form $A e^{-B}$.  To
leading approximation, $B$ is the bounce action obtained by
replacing the tree-level potential by the leading one-loop
approximation to the effective potential, in agreement with the
generally adopted {\it ad hoc} remedy.  The next correction to $B$
(which is proportional to an inverse power of a small coupling) is
given in terms of the next-to-leading term in the effective
potential and the leading correction to the two-derivative term in
the effective action.  The corrections beyond these (which may be
included in the prefactor) do not have simple expressions in terms of
the effective potential and the other functions in the effective
action.  In particular, the scalar-loop terms which give an
imaginary part to the effective potential do not explicitly appear;
the corresponding effects are included in a functional determinant
which gives a manifestly real result for the nucleation rate.

\vskip .1in
\noindent\footnote{}{\twelvepoint This work was supported in part by the US
Department of
Energy and by the National Science Foundation under Grant No. PHY89-04035}

\vfill\eject

\baselineskip=20pt
\pagenumbers
\pageno=1

\def\half{{ {1\over 2}}}
\def\tr{\mathop{\rm tr}}
\def\seff{S_{\rm eff}}
\def\veff{V_{\rm eff}}
\def\fv{{\rm fv}}
\def\b{{\rm b}}
\def\ct{{\rm ct}}

\def\vh{\hat V}
\def\zh{\hat Z}
\def\bphi{\bar\phi}

\def\sqr#1#2{{\vcenter{\hrule height.#2pt
        \hbox{\vrule width.#2pt height#1pt \kern#1pt
                \vrule width.#2pt}
        \hrule height.#2pt}}}
\def\square{\mathchoice\sqr64\sqr64\sqr{2.1}3\sqr{1.5}3}

\catcode`@=11
\def\@versim#1#2{\lower.7\p@\vbox{\baselineskip\z@skip\lineskip-.5\p@
    \ialign{$\m@th#1\hfil##\hfil$\crcr#2\crcr\sim\crcr}}}
\def\simge{\mathrel{\mathpalette\@versim>}} %
\def\simle{\mathrel{\mathpalette\@versim<}} %
\catcode`@=12 

\chapter{ Introduction }

     When studying quantum field theories in a cosmological context
one often encounters situations where the field is not at the
absolute minimum of the potential (the ``true vacuum"), but is
instead at some local minimum (a ``false vacuum") which is higher
in energy.  The transition to the true vacuum state proceeds by a
quantum mechanical tunneling process in which the field in a
small region of space tunnels through the potential energy
barrier to form a bubble of true vacuum. Once nucleated, the
bubble expands, converting false vacuum to true as it does so.

     The bubble nucleation rate per unit volume, $\Gamma$, can
be calculated by a method, due to Coleman\rlap,
\Ref\COLEMAN{S.~Coleman, \pr{15}{2929}77.  } which is based on
finding a ``bounce" solution to the
classical Euclidean field equations.  Thus, for a theory with a
single scalar field, one must solve
$$  \partial_\mu\partial_\mu\phi \equiv \square \phi
    = {\partial V\over \partial\phi}
           \eqn\bounceeq  $$
subject to the boundary condition that $\phi$ approach its false
vacuum value as any of the $x_\mu$ tend to $\pm\infty$.  $\Gamma$
can then be written in the form
$$  \Gamma = A\,e^{-B}
      \eqn\aexpb $$
where $B$ is the Euclidean action of the bounce, while $A$ is
given by an expression involving functional determinants. In
practice, the latter generally turns out to be quite difficult to
evaluate, although one can show that $A$ is equal to a
numerical factor of order unity times a dimensionful quantity
determined by the characteristic mass scales of the theory.

     A difficulty arises when one deals with theories where the
symmetry breaking is a result of radiative corrections\rlap.
\Ref\COLEMANWEINBERG{S.~Coleman and E.~J.~Weinberg, \pr7{1888}73.}  In  such
cases the true vacuum is not determined by $V(\phi)$, but instead
 can be found only by examining the effective potential,
$\veff(\phi)$.   The bounce equation \bounceeq\ is clearly
inappropriate --- in fact,  if $V(\phi)$ has only a single
minimum there will not be any bounce solution.  An obvious
alternative, which has been taken by a number of authors\rlap,
\ref{For some early examples, see P.~Frampton, \prl{37}{1378}76 and
\pr{15}{2722}77; A.~Linde, \pl{70}{306}77.}
is to modify this equation by
substituting $\veff(\phi)$ for $V(\phi)$.

     Although plausible, and clearly a step in the right direction,
this procedure raises some questions.  The one-loop radiative
corrections generate an effective action which contains not only
$\veff$, but also terms involving derivatives (of all orders) of the
fields.  Can these terms be neglected when dealing with
configurations, such as the bounce solution, which are not constant in
(Euclidean) space-time?  Even if this can be done in a first
approximation, what are the nature and magnitude of the corrections
which these terms generate?  There are also questions relating to
$\veff$ itself.  First, the effective potential obtained by
perturbative calculations differs considerably from that defined by a
Legendre transform (the latter must be convex, while the former is
not).  The latter clearly does not lead to an appropriate bounce, but
how precisely does the formalism pick out the former?  Further, the
perturbative effective potential is known to be complex for certain
values of the fields.  How is the imaginary part of
$\veff$ to be handled?

     In order to answer these questions I develop in this paper a
scheme for calculating the bubble nucleation rate in theories with
symmetry breaking by radiative corrections.  The general idea is to
use the path integral approach of Callan and Coleman\rlap,
\Ref\CALLANCOLEMAN{C.~G.~Callan and S.~Coleman, \pr{16}{1762}77.} but
to integrate
out certain fields at the outset.  This leads to a modified effective
action which, although it differs somewhat from the usual perturbative
$\seff$, gives a correct description of the vacuum structure of the
theory and has a bounce solution which can provide the basis for a
tunneling calculation.  The final result for $\Gamma$ may again be
written in the form of Eq.~\aexpb, with $A$ either of order unity
or proportional to a small inverse power of the coupling.  As might
be expected, the leading approximation to $B$ is obtained simply by
replacing $V$ with the dominant one-loop part of $\veff$ in the
standard procedure.  The next-to-leading terms in $\veff$, as well as
the first corrections to the derivative terms in the effective action,
then come in and give calculable and significant (i.e., larger than
order unity) corrections to $B$.  However, the corrections beyond this
point do not have a simple expression in terms of $\veff$ and the
other functions entering $\seff$.  In particular, the potentially
complex parts of the effective potential do not enter the calculation
directly, but only as part of more complicated functional determinants
which can easily be shown to be real.

     This approach can be understood by recalling the
Born-Oppenheimer approximation for the calculation of molecular energy
levels.  Because the natural time scale for the electrons is much shorter
than that for the nuclei, the electrons can be treated as adapting almost
instantaneously to changes in the positions of the nuclei.  This leads to
an effective action, involving only the nuclear coordinates, which may be
used to describe situations in which the nuclei do in fact move slowly;
thus, it can be applied to the vibrational spectrum of the molecule,
but should not be used when studying the scattering of high energy
particles off the nucleus.

     A similar situation arises in theories where radiative corrections
change the vacuum structure. For definiteness, consider the case of scalar
electrodynamics\rlap,\refmark{\COLEMANWEINBERG} where radiative
corrections from photon-loop diagrams lead to spontaneous symmetry
breaking if the scalar field self-coupling is $O(e^4)$ and the scalar mass
term is sufficiently small.  This relationship between the couplings
implies that the natural time scale for the variation of the
electromagnetic field is much less than that for the scalar field.
Consequently, if we are only interested in the long wavelength modes of
the latter, as is the case when studying vacuum tunneling, we can
integrate out the electromagnetic field to obtain an effective action for
the scalar field.

    The remainder of this paper is organized as follows.  Sec.~II
reviews the method of Callan and Coleman.  In Sec.~III, I show how this
method can be adapted to theories with radiative symmetry breaking.  The
treatment here is somewhat formal, so that it can be applied to a
variety of theories.  The concrete implementation of the method in
specific models is developed in Secs.~IV and V, which discuss a
simple model with two scalar fields and scalar electrodynamics,
respectively.  Sec.~VI contains some concluding remarks.

\vfill
\eject

\chapter{The Callan-Coleman Formalism}

     Callan and Coleman\refmark{\CALLANCOLEMAN} evaluated $\Gamma$ by
calculating the imaginary part of the energy of the false vacuum.  This
can  be obtained from the quantity
$$ \eqalign {G(T)&= \langle \phi({\bf x})=\phi_{\fv}|\,e^{-HT}\,|
   \phi({\bf x})=\phi_{\fv} \rangle \cr
           &=\int [d\phi]e^{-[S(\phi) +S_{\ct}(\phi)] }\cr}
    \eqn\GofT  $$
where the Euclidean action
$$ S(\phi) =\int d^3x \int^{T/2}_{-T/2}dx_4
   \left[{1 \over 2} (\partial_\mu\phi)^2 + V(\phi)\right]
       \eqno\eq $$
is expressed in terms of renormalized fields and parameters,
while $S_{\ct}$ contains the counterterms needed to make the
theory finite.  Although divergent, these counterterms are of
higher order in the coupling constants and, as usual, are treated
as perturbations.

    The path integral is over all configurations such that
$\phi$ takes its false vacuum value $\phi_{\fv}$ at $\pm T/2$ and
at spatial infinity. In the limit $T \rightarrow \infty $
Eq.~\GofT\ is dominated by the lowest energy state with a
non-vanishing contribution (i.e., the false vacuum) and is of the
form
$$ G(T) \approx e^{-{\cal E} T\Omega}
   \eqno\eq $$
where $\Omega$ is the volume of space and $\cal E$ may be
interpreted as the energy density of the false vacuum state.  Because
this is an unstable state, $\cal E$ is complex, with its
imaginary part giving the decay rate, which in this case is
simply the bubble nucleation rate.  Dividing by $\Omega$ gives
the nucleation rate per unit volume,
$$  \Gamma = -2 {\rm Im}\, {\cal E}
    \eqno\eq $$

     The path integral may be approximated as the sum of the
contributions about all of the stationary (or quasi-stationary)
points of the Euclidean action $S(\phi)$: the pure false vacuum,
the bounce solution with all possible locations in Euclidean
space-time, and all multibounce configurations. In each case the
contribution to the path integral is obtained by expanding the
field about the classical solution $\bphi(x)$:
$$ \phi(x) = \bphi(x) + \eta(x)
  \eqno\eq$$
and then integrating over $\eta$.  To leading approximation one
keeps only the terms in the action which are quadratic in $\eta$.
Expanding these in terms of the normal modes of $ S''(\bphi) =
-\square + V''(\bphi) $ gives a product of Gaussian integrals.
The terms of cubic and higher order in $\eta$
can then be treated as perturbations, and
have the effect of multiplying the Gaussian approximation for
$\Gamma$ by a power series in the coupling constants, with the
first term being unity.  The counterterms are also treated as a
perturbation, with the leading contribution being from $S_{\ct}(\bphi)$.

     Thus, the integration about the trivial solution $\phi(x) =
\phi_{\fv}$ yields
$$ \eqalign{ G_0 &=  NK_0\, e^{- S(\phi_{\fv})}
   e^{-S_{\ct}(\phi_{\fv})} (1 + \cdots ) \cr
     &=  NK_0\, e^{-\Omega TV(\phi_{\fv})}
   e^{-S_{\ct}(\phi_{\fv})} (1 + \cdots ) }
   \eqno\eq$$
where $N$ is a normalization factor,
$$ \eqalign {K_0 &= [\det S''(\phi_{\fv})]^{-1/2}\cr
  &= [\det (- \square + V''(\phi_{\fv}))]^{-1/2}\cr}
  \eqno\eq$$
and the dots represent terms, due to higher order
perturbative effects, that can be neglected at the order to which we
are working.  Although the determinant is actually divergent, the
divergences are cancelled by the factor of $e^{-S_{\ct}} $.

    The calculation of the contribution from the bounce solution
$\phi_\b(x)$ is similar, but must be modified to take into account
the fact that $S''(\phi_\b)$ has one negative eigenvalue and four
zero eigenvalues; for none of the corresponding modes is the
integral truly Gaussian.  The zero-frequency modes are treated by
introducing collective coordinates corresponding to the position
of the bounce.  Integrating over these gives a factor of $\Omega
T$ and introduces a Jacobean factor which, for $O(4)$-symmetric
bounces, is given by
$$  J = {B^2\over 4\pi^2}
    \eqno\eq $$
The negative mode is handled by deforming the contour of
integration.  Aside from a factor of $1/2$, this leads to a
contribution whose imaginary part is just that which would have
been obtained from a naive application of the Gaussian integration
formulas; i.e.,
$$ {\rm Im} \, G_1 = G_0 G_\b
   \eqno\eq $$
where
$$  G_\b = K_1 J \Omega T\, e^{- B}
    e^{-[S_{\ct}(\phi_\b) - S_{\ct}(\phi_{\fv})] }
    \, (1 + \cdots)
    \eqno\eq $$
with
$$ B= S(\phi_\b) - S(\phi_{\fv})
    \eqno\eq$$
and
$$ \eqalign{ K_1 &= {1\over 2} K_0^{-1}
   \left| {\det}'[- \square + V''(\phi_\b)] \right|^{-1/2} \cr
    &= {1\over 2} \left|{ {\det}'[- \square + V''(\phi_\b)] \over
     {\det}[- \square + V''(\phi_{\fv})]} \right|^{-1/2} }
 \eqno\eq$$
Here $\det '$ indicates that the translational zero-frequency
modes are to be omitted when evaluating the determinant.  As
before, the divergences in the determinant factors are cancelled
by the terms containing $S_{\ct}$.

     Finally, the $n$-bounce quasi-stationary points give a
contribution of the form $G_n = G_0 G_\b^n /n! $.  The bounce
contributions then exponentiate, and one finds that
$$ \eqalign{ \Gamma & = 2 {G_\b \over \Omega T} \cr
 &=   {B^2 \over 4\pi^2}\,   e^{- B}
     \left|{ {\det}'[- \square + V''(\phi_\b)] \over
     {\det}[- \square + V''(\phi_{\fv})]} \right|^{-1/2}
    e^{-[S_{\ct}(\phi_\b) - S_{\ct}(\phi_{\fv})] }
      \, (1 + \cdots)  }
   \eqn\oldanswer$$

     For comparison with later results we need to know the order
of magnitude of the various terms in this expression.  To be
specific, let us assume that we can identify a small coupling
$\lambda$ and a dimensionful quantity $\sigma$ such that
the potential can be written in the form
$$ V(\phi) = \lambda \sigma^4 U(\psi)
    \eqno\eq $$
where $U$ involves no small couplings and the dimensionless field
$ \psi = {\phi /\sigma} $.
The minima of the potential must then be located either at
$\phi=0$ or at values of $\phi$ of order $\sigma$.
(With a single scalar field the most general renormalizable
potential is of the form
$$ V(\phi) = {m^2\over 2} \phi^2 + {c\over 3}\phi^3 +
    {\lambda\over 4} \phi^4
   \eqno\eq$$
The above assumption is then equivalent to assuming that $ c $ is
of the order of $ \lambda \, \sigma$, with $\sigma \equiv
m/\sqrt{\lambda} $. )

     By defining a dimensionless variable $ s = \sqrt{\lambda}\,
\sigma x$, we may write the field equations as
$$    \square_s \psi = {\partial U\over \partial \psi}
      \eqno\eq $$
{}From the assumptions made above, this equation involves no small
parameters and so has a bounce solution in which $\psi$ is
of order unity and differs from the false vacuum within a
region of a spatial extent (measured in terms of
$s$) which is also of order unity.  In terms of the original
variables, the bounce has $\phi$ of order $\sigma$ and
extends over a range of $x$ of order $1/(\sqrt{\lambda}
\,\sigma)$.

    With the same change of variables, the action becomes
$$  S = {1 \over \lambda} \int d^4s\,
  \left[{1\over 2} \left({\partial\psi \over \partial s^\mu}\right)^2
         + U(\psi)\right]
    \eqno\eq $$
Since the integrand contains no small parameters, while the
volume of the bounce restricts the integration to a region of
order unity, the bounce action $B$ is of order $\lambda^{-1}$.

    Similarly, the determinant factor $K_1$
becomes
$$ \eqalign {K_1 &= {1\over 2}
     \left|{ {\det}'[(\lambda\sigma^2) (- \square_s +
U''(\psi_\b))] \over
     {\det}[(\lambda\sigma^2) (- \square_s + U''(\psi_{\fv}))]}
      \right|^{-1/2} \cr
    &=  {1\over 2} \lambda^2\sigma^4
     \left|{ {\det}'[- \square_s + U''(\psi_\b)] \over
     {\det}[- \square_s + U''(\psi_{\fv})]} \right|^{-1/2} }
 \eqno\eq$$
where the explicit factor of $\lambda^2\sigma^4$ on the second line
arises because the ${\det}'$ factor involves four fewer modes than the
$\det$ factor.  With this factor extracted, the ratio of determinants
is formally of order unity, although divergent.\footnote{1}{It was
asserted above that these divergences are cancelled by divergences in
$S_{\ct}(\bphi) - S_{\ct}(\phi_{\fv})$.  To see that these counterterms
are of the correct order of magnitude, note that with our assumptions
the one-loop contributions to the counterterm Lagrangian are of order
$\lambda^2\sigma^4$; since the bounce has a spatial volume of order
$1/(\sqrt{\lambda}\sigma)^{4}$, the difference in the counterterm
actions is of order unity, which is what is required.} Finally, the
Jacobean factor $J$ is proportional to $B^2 \sim
\lambda^{-2}$.  Putting all of these factors together, we see
that the nucleation rate is of the form
$$  \Gamma = c_1 \sigma^4\, e^{c_2/\lambda}
    \eqno\eq$$
with $c_1$ and $c_2$ both of order unity.

\vfill
\eject

\chapter{A Formalism for Theories with Radiative Symmetry Breaking }

     I now describe how the methods of the previous section can
be adapted to theories where radiative corrections to the tree-level
potential qualitatively change the vacuum structure.  The general
formalism is derived in this section, while the detailed
implementation for two specific theories is described in Secs.~IV and V.

     In theories of this sort the fields may be divided into two sets,
denoted $\chi_j$ and $\phi_j$.  The former give rise to the radiative
corrections which effect the changes in the vacuum structure, while
the various vacua are distinguished by the value of the latter.  In
order that the radiative corrections from the $\chi$-loops be
comparable to the tree-level terms, the interactions among the
$\phi_j$ must be weak compared to the $\phi$-$\chi$ interactions.
Hence, the $\chi_j$ may be viewed as ``fast'' degrees of freedom which
can be integrated out in a Born-Oppenheimer type approximation.  For
simplicity of notation, I will assume here a theory with only field of
each type, both assumed to be scalars; the generalization to more
complicated theories is straightforward.

     As usual, the bubble nucleation rate is obtained from
the quantity
$$  G(T)= \int [d\phi][d\chi] e^{-S(\phi,\chi)}
        \eqno\eq$$
(To simplify the equations the counterterm action has not been
explicitly written.)  Because $S(\phi,\chi)$ does not
display the correct vacuum structure, its classical field
equations do not have a bounce solution, and so the methods of
Sec.~II cannot be directly applied.  This difficulty can be
circumvented by integrating over $\chi$ to obtain
$$ G(T) = \int [d\phi] e^{-W(\phi)}
       \eqno\eq$$
where
$$ W(\phi) = -\ln\, \int [d\chi] e^{-S(\phi,\chi)}
        \eqn\wdef $$
may be thought of as an effective action.  $W$ has a simple graphical
interpretation.  The integral on the right hand side of Eq.~\wdef\ is
equal to the sum of all vacuum Feynman diagrams in a theory of a
quantized $\chi$-field interacting with a fixed c-number $\phi$-field;
i.e., the sum of all graphs with only internal $\chi$-lines and
external $\phi$-lines.  Its logarithm gives the sum
of all connected vacuum graphs.  In general these include
one-particle reducible graphs, although in many cases these are
eliminated by symmetry considerations.

    $W(\phi)$ should be compared with the more familiar effective
action, $S_{\rm eff}(\phi, \chi)$, which generates one-particle
irreducible Green's functions.  For theories in which $W(\phi)$
only receives contributions from one-particle irreducible graphs,
$S_{\rm eff}(\phi, 0)$ and $W(\phi)$ differ by the contribution from
graphs with internal $\phi$-lines.  Our assumptions about the
size of the $\phi$ self-couplings imply that these graphs are
suppressed.  Hence, the dominant terms in $S_{\rm eff}(\phi, 0)$ and
$W(\phi)$ agree, and so the latter correctly reflects the vacuum
structure of the theory.

    One might therefore envision using the classical
field equations implied by $W$ to determine a bounce solution
which would be the basis for a nucleation rate calculation. Two
practical difficulties arise. First, one cannot in general obtain
a closed form expression for $W$, but only a perturbative series.
Second, $W(\phi)$ is a nonlocal functional, so that even if a
closed-form expression were available, the equations determining
its stationary points would be rather unpleasant.

    One way around these difficulties is to find a local action
$W_0(\phi)$ which is a sufficiently close approximation to $W$ and
then use its stationary points as the basis for the nucleation rate
calculation.  The possibility of such an approximation arises from the
fact that the nonlocality of $W$ is significant only on distances
shorter than a characteristic size set by the $\chi$-field
interactions.  For $\phi$-fields which are slowly varying relative to
these scales, $W(\phi)$ can be expanded in a ``derivative expansion''
of the form
$$ W(\phi) = \int d^4x \left[ \vh(\phi)
    + \half \zh(\phi) (\partial_\mu\phi)^2
      +   \cdots \right]
           \eqno\eq $$
where the dots represent terms with four or more
derivatives.  (A similar expansion of the effective action,
$$ S_{\rm eff}(\phi,\chi) = \int d^4x \left[ V_{\rm eff}(\phi,\chi)
    + \half Z_\phi(\phi,\chi) (\partial_\mu\phi)^2
    + \half Z_\chi(\phi,\chi) (\partial_\mu\chi)^2
      +   \cdots \right]
           \eqno\eq $$
is often made.  In line with the previous remarks, $\vh(\phi)$
and $\zh(\phi)$ differ from $V_{\rm eff}(\phi,0)$ and
$Z_\phi(\phi,0)$ by the omission of graphs with
internal $\phi$-lines.)

     The desired approximate
action is obtained by keeping only the first two terms in the
derivative expansion of $W(\phi)$,
and then only the lowest order contributions to these; thus
$$  W_0(\phi) = \int d^4x \left[ {1\over 2} (\partial_\mu \phi)^2
   + \hat V_{\rm 1-loop}(\phi) \right]
     \eqn\wzero $$
Here $\vh_{\rm 1-loop}(\phi)$ is the sum of the
tree-level $V(\phi)$ and the contributions from graphs with a
single $\chi$-loop; to this order it is equal to
$V_{\rm eff}(\phi,0)$.  Since the loop corrections included in
$\vh_{\rm 1-loop}(\phi)$ are precisely those responsible for altering
the vacuum structure of the theory, $W_0$ possesses an
appropriate bounce solution.

     Thus, for slowly varying $\phi$ we may write
$$  W(\phi) = W_0(\phi) + \delta W(\phi)
     \eqno \eq $$
with $\delta W(\phi)$ representing subdominant terms.  We can (and
will) make the same decomposition for arbitrary $\phi$, but
$\delta W$ will  then not necessarily be small. In particular,
because the path integral includes contributions from fields with
Fourier components of arbitrarily high momentum, for which the
derivative expansion is not valid, $\delta W$ cannot be treated as a
small perturbation within the path integral.

     Nevertheless, the path integral may be approximated by
expanding about the stationary points of $W_0$, provided that
these are themselves slowly varying functions.  Thus, let us
suppose that the field equation implied by $W_0$,
$$   \square \phi = {\partial\vh_{\rm 1-loop} \over\partial \phi}
    \eqno\eq$$
has a solution $\bphi(x)$.  Expanding $W$ about this solution gives
$$  W(\phi) = W(\bphi) + \int d^4z \, W'(\bphi;z) \eta(z)
     + {1\over 2} \int d^4z\, d^4z' \, W''(\bphi;z,z')\eta(z) \eta(z')
     + O(\eta^3)
     \eqn\etaexp $$
where $\eta(x) = \phi(x) - \bphi(x)$, and primes denote
variational derivatives; e.g.,
$$  W'(\bphi; z) \equiv
  \left.{\delta W(\phi) \over \delta \phi(z)} \right|_{\phi=\bphi}
     \eqno \eq $$
(The term linear in $\eta$ is nonvanishing because $\bphi$ is not
a stationary point of the full $W(\phi)$.)

     We now insert Eq.~\etaexp\ into the path integral.  The cubic
and higher order terms in $\eta$ can be treated as small
perturbations.  If these are dropped, the integral becomes
Gaussian.  About the false vacuum, $\bphi(x) = \phi_{\fv}$, $W''(\bphi)$
has no zero or negative eigenvalues and the integration gives
$$  G_0 = e^{-W(\phi_{\fv})} \,
      e^{  W'(\phi_{\fv}) [W''(\phi_{\fv})]^{-1} W'(\phi_{\fv}) } \,
      [\det W''(\phi_{\fv})]^{-1/2}
     \eqno \eq $$
(For the sake of compactness, the spatial arguments of the variational
derivatives and the associated spatial integrations have not been
explicitly displayed.)
As in the standard case, the bounce solution has
zero frequency modes which must be handled separately.
\footnote{2}{Because
the bounce is obtained from the truncated action $W_0$,
the usual translational modes, proportional to $\partial_\mu\phi_\b$,
differ slightly from the zero modes of $W''$; however, it can be shown
that the effects due to the difference are suppressed by a power of
the coupling constant and hence can be
neglected at the order to which we will be working.}
Introducing collective coordinates and proceeding as usual leads to
$$  Im G_1  = {1\over 2} \Omega T\, e^{-W(\phi_\b)} \,
      e^{  W'(\phi_\b) [W''(\phi_\b)]^{-1} W'(\phi_\b) } \,
      |{\det}' W''(\phi_\b)|^{-1/2}\, J
     \eqno \eq $$
where now ${\det}'$ indicates that the determinant is restricted
to the subspace orthogonal to the zero modes of $W_0''$.
Similarly, $(W'')^{-1}$ is to be evaluated in this subspace,
while the Jacobean $J$ is given by
$$  J = { [ W_0(\phi_\b) - W_0(\phi_{\fv})]^2 \over 4\pi^2 }
      \eqno\eq $$

    The path from these results to the nucleation rate is just
as before, and gives
$$  \Gamma = e^{-C_1} \, e^{C_2}
  \left| {{\det}' W''(\phi_\b) \over {\det} W''(\phi_{\fv}) }
      \right|^{-1/2} \, J  (1 + \cdots)
    \eqn\gammaexp $$
where
$$  C_1 = W(\phi_\b) - W(\phi_{\fv})
  \eqno\eq $$
and
$$  C_2 = W'(\phi_\b) [W''(\phi_\b)]^{-1} W'(\phi_\b)
       - (\phi_\b \rightarrow \phi_\fv)
  \eqn\ctwo$$

      The next step is to expand the various terms in Eq.~\gammaexp\
in powers of the coupling constants.  The details of this depend on
the structure of the theory and the magnitudes of the various
couplings.  The general features described below hold for the specific
theories considered in the next two sections and are likely to be true
for all theories with radiative symmetry breaking. As in Sec.~II, the
expansion of $\Gamma$ is only carried out to terms of order unity;
terms which do not contribute to that order are represented by dots.

    First, the bounce solution has a characteristic length scale which
is large compared to those characterizing the $\chi$-$\chi$ and
$\chi$-$\phi$ interactions.  It therefore varies slowly enough to allow
a derivative expansion of $W(\phi_\b)$.  (We will see that in some
situations this expansion must be terminated after the first few
terms, leaving a nonlocal remainder.  This does not materially affect
the picture outlined here.)  For both  $\bar\phi=\phi_\b$
and $\bar\phi=\phi_{\fv}$, it is then possible to write $W(\bar\phi)$
as a sum of terms of successively higher order in the couplings:
$$  W(\bphi)  = W_0(\bphi) + W_1(\bphi) + W_2(\bphi) +  \cdots
    \eqn\woneetc $$
with $W_0$ given by Eq.~\wzero\ and the remaining terms containing
higher order contributions to the derivative expansion.  For example,
in the theories considered below $W_1$ contains two-loop contributions
to $\vh$ and one-loop contributions to $\zh$.

     A similar expansion can be
used for $W'(\bphi;z)$.   The zeroth order term vanishes because
$\bphi$ is a stationary point of $W_0$, and we have
$$  W'(\bphi;z)  =  W'_1(\bphi;z)  + \cdots
    \eqn\wprime $$
Matters are less simple for $W''(\bphi;z,z')$. Although the
derivative expansion can be used when $|z-z'|$ is large, the
behavior for small $|z-z'|$ is sensitive to the high momentum
modes and so the derivative expansion fails even for constant
$\bphi$.   However, the relation
$$  W'' = W_0'' \left[ 1 + (W_0'')^{-1} \delta W'' \right]
\eqn\wprimeprimepert $$ can be used to obtain formal expansions for
$(W'')^{-1}$ and  $\det W''$ as power series in $(W_0'')^{-1} \delta
W''$.   The actual utility of these expansions depends on the size of
the contribution from the region of small $|z-z'|$. In the
calculation of $C_2$ this contribution is subdominant and
$$  C_2 = W'_1(\phi_\b) [W''_0(\phi_\b)]^{-1} W'_1(\phi_\b)
  - (\phi_\b \rightarrow \phi_{\fv} ) + \cdots
    \eqno\eq $$
For the determinant factor, on the other hand, more terms must be
retained:
$$ \eqalign{  {\det}'[W''] &= {\det}'[W_0'']
   \det [ I + (W_0'')^{-1} \delta W''] \cr
    &= {\det}'[W_0'']
    \exp\left\{\tr \ln [ I + (W_0'')^{-1} \delta W''] \right\} \cr
     &= {\det}'[W_0'']
    \exp\left\{\tr (W_0'')^{-1} \delta W''
      + {1\over 2} \tr \left[(W_0'')^{-1} \delta W''\right]^2
                + \cdots\right\} }
        \eqn\detexp $$

        Inserting these expansions into Eq.~\gammaexp\ leads to
$$  \Gamma = e^{-B_0} \, e^{-B_1} \, e^{-B_2}
    \left| {{\det}' W_0''(\phi_\b) \over {\det} W_0''(\phi_{\fv}) }
      \right|^{-1/2} \, {B^2_0 \over 4\pi^2} \, (1 + \cdots)
       \eqn\finalexp $$
where
$$  B_0 = W_0(\phi_\b) - W_0(\phi_{\fv})
       \eqno\eq $$
$$  B_1 = W_1(\phi_\b) + {1\over 2}\tr [W_0''(\phi_\b)]^{-1} \delta
W''(\phi_\b)
   - (\phi_\b \rightarrow \phi_{\fv} )
       \eqno\eq $$
and
$$ \eqalign{  B_2 &= W_2(\phi_\b)
  -  W'_0(\phi_\b) [W''_0(\phi_\b)]^{-1} W'_0(\phi_\b)
   + {1\over 4} \tr \left\{[W_0''(\phi_\b)]^{-1}
    \delta W''(\phi_\b) \right\}^2  \cr
    & \qquad   - (\phi_\b \rightarrow \phi_{\fv} )}
       \eqno\eq $$
We thus have the naive expression for $\Gamma$, obtained by
the standard procedure  with $V(\phi)$ replaced by the one-loop
effective potential, multiplied by a correction factor
$e^{-(B_1 + B_2)}$.  We will find that $B_1$ is proportional to
an inverse power of the couplings, while $B_2$ is of order unity.
Thus, although the $B_2$ correction is of the same order of
magnitude as the determinant factor, which in practice cannot be
evaluated precisely, the $B_1$ factor modifies the naive result
in a significant and (at least numerically) calculable manner.

     The form of these correction factors can be understood. $B_0$ is
obtained by evaluating the approximate action $W_0$ at its stationary
points, whereas what we really want is the exact action $W$ evaluated
at its own stationary points.  The difference between these two
quantities can be obtained by treating $\delta W = W_1 + W_2 + \cdots$
as a small perturbation; up to second order, this gives the first term
in $B_1$ and the first two terms in $B_2$.  As we will see below, the
remaining terms correct for the fact that $W$ is not the full
effective action and take into account contributions to the full
effective action which would otherwise be neglected.

\vfill\eject

\chapter{Example 1: A Scalar Field Theory}

    As a first example, consider a theory with two scalar fields
governed by the Lagrangian
$$    {\cal L} = {1\over 2} \left(\partial_\mu\phi\right)^2
       + {1\over 2} \left(\partial_\mu\chi\right)^2
        -V(\phi,\chi) + {\cal L}_{\ct}
     \eqno\eq$$
with
$$  V = {1\over 2} \mu^2 \phi^2 + {\lambda\over 4!} \phi^4
    + {1\over 2} m^2 \chi^2  +  {f\over 4!} \chi^4
       + {1\over 2}g^2\phi^2\chi^2
     \eqno\eq $$
Both $\mu^2$ and $m^2$ are positive. In order that the $\chi$-loop
corrections to the effective potential be able to generate a
symmetry-breaking minimum at a value $\phi = \sigma \ne 0$, $\lambda$
must be $O(g^4)$, while $f$ should be $O(g^2)$ to ensure that the
$\chi \rightarrow - \chi$ symmetry remains unbroken.  There is also an
upper bound of order $g^2\sigma$ on $\mu$, which will be displayed in
detail below.  Finally, it will be convenient to assume that $m$ is
$O(g\sigma)$.

     Renormalization conditions must be specified in order to fix the
counterterms in ${\cal L}_{\ct}$.   For the lowest order approximation
to the effective potential only the definitions of $\mu^2$ and
$\lambda$ are needed.   These can be given by choosing an arbitrary
renormalization scale $\tilde\sigma$ and requiring that
$$  \mu^2 = \left.{\partial^2 V_{\rm eff} \over \partial \phi^2}
   \right|_{\phi=\chi=0}
   \eqn\mufix $$
and
$$  \lambda = \left.{\partial^4 V_{\rm eff} \over \partial \phi^4}
   \right|_{\phi=\tilde \sigma, \chi=0}
   \eqn\lambdafix $$

    For $\chi=0$ the effective potential is given by the sum of graphs
with only external $\phi$-lines, all carrying zero four-momentum. The
leading contributions are from tree graphs and graphs with one
$\chi$-loop, as well as the corresponding counterterms.
Let us denote  \footnote{3}{Since $\chi=0$ is to be understood for the
remainder of this discussion, $\veff$ and related quantities will be
written as functions of a single variable.} the sum of these by $V_{\rm
1-loop}(\phi)$, although it should be kept in mind that this does not
include the graphs with one $\phi$-loop. We then have, up to an
additive constant,
$$ \eqalign {V_{\rm 1-loop}(\phi) &=
      {1\over 64\pi^2} (m^2 + g^2 \phi^2)^2
   \left[ \ln  \left( {m^2 + g^2 \phi^2
  \over m^2 + g^2 \tilde\sigma^2 } \right) - {1\over 2} \right] \cr
  &  -{1\over 4}\left({\mu^2 \over \tilde\sigma^2} + y_1\right)
      \left( \phi^2 - \tilde\sigma^2 \right)^2
   + \left[ {1\over 4!} (\lambda + z_1) + {1\over 4}
  \left({\mu^2 \over \tilde\sigma^2}+ y_1\right) \right] \phi^4 }
     \eqno\eq  $$
where
$$ y_1 = {g^2m^2\over 16\pi^2} \ln \left( {m^2 + g^2 \tilde\sigma^2
        \over m^2 }\right)
     \eqno\eq $$
and
$$ z_1 = {g^4 \over 8\pi^2}
 \left[ {4g^4 \tilde\sigma^4 \over (m^2 + g^2 \tilde\sigma^2)^2}
    - {12 g^2 \tilde\sigma^2 \over m^2 + g^2 \tilde\sigma^2} -3
\right]
     \eqno\eq $$
It is most convenient to choose the renormalization point
$\tilde\sigma$ to be $\sigma$, the location of the minimum of
$V_{\rm 1-loop}$.  The requirement $V'_{\rm 1-loop}(\sigma) =0$ then
relates $\lambda$ to the other parameters via
$$ \lambda = - 3\left({\mu^2 \over \sigma^2} + y_1\right) -z_1
    \eqno\eq $$
and yields
$$ \eqalign {V_{\rm 1-loop}(\phi) &=
   {1\over 64\pi^2 }(m^2 + g^2 \phi^2)^2
   \left[ \ln  \left( {m^2 + g^2 \phi^2
  \over m^2 + g^2 \sigma^2 } \right) - {1\over 2} \right] \cr
    &\qquad   -{1\over 4}\left({\mu^2 \over \sigma^2} + y_1\right)
      \left( \phi^2 - \sigma^2\right)^2 }
  \eqn\voneloop  $$
For $\sigma$ to be a minimum, $V_{\rm 1-loop}''(\sigma)$ must be
positive, which leads to the bound
$$  \mu^2 < {g^4\sigma^2 \over 16\pi^2} \left[ 4 -
    {m^2 \over g^2\sigma^2} \ln \left( {m^2 + g^2 \sigma^2
        \over m^2} \right) \right]
       \eqno\eq $$
A slightly stronger bound on $\mu$ is obtained if one requires,
as we will, that the minimum at $\phi=\sigma$ be deeper than that
at $\phi=0$, so that the latter corresponds to a
metastable false vacuum.

     The higher order contributions to the effective potential can
be obtained from Feynman graphs with ``dressed'' propagators that
take into account the background $\phi$-field.  Thus, the
effect of summing over all numbers of interactions with a
constant external $\phi$-field is to replace the
standard $\chi$-propagator $(k^2 - m^2)^{-1}$ by
$$   [k^2 - m^2 - g^2\phi^2] ^{-1}
     \equiv [k^2 - M^2(\phi)] ^{-1}
     \eqno\eq$$
Because the $\chi$-loop
contributions to $V_{\rm eff}$ are the of the same magnitude as the
tree terms, the dressed $\phi$-propagator is somewhat more
complicated.  In order that the perturbative order of a graph
increase with the number of loops, these one-loop terms must be
included in the dressed propagator, which then takes the form
$$   [k^2 - V''_{\rm 1-loop}(\phi) ] ^{-1}
     \equiv [k^2 - {\cal M}^2(\phi)] ^{-1}
     \eqno\eq$$
This inclusion of loops in the propagator means that the loop-counting
for a graph with dressed propagators will not necessarily agree with
that for the corresponding graphs with elementary propagators.   Thus,
the one $\phi$-loop contribution to the effective potential will
include graphs which are multiloop when drawn with elementary
propagators.  Some examples of these are shown in Fig.~1.
\FIG\one{Examples of graphs, which although multiloop when drawn in
terms of elementary propagators, are included in the one $\phi$-loop
contribution to the effective potential in the model of Sec.~4.
Solid and wiggly lines
represent $\phi$ and $\chi$ propagators, respectively.}
In the
last of these the momentum assignments have been shown explicitly in
order to emphasize that they are not the ones usually associated with
this graph.   It should be clear from this example that graphs with
both $\chi$ and $\phi$ internal lines must be analyzed carefully to
avoid double-counting.

     The next contributions to $V_{\rm eff}$, of order $fg^4$ and
$g^6$, arise from the two-loop graphs in Fig.~2,
\FIG\two{The two-loop graphs which contribute to the effective
potential in the model of Sec.~4 at order $fg^4$ and $g^6$.}
the one-loop counterterm graphs in Fig.~3,
\FIG\three{The one-loop counterterm graphs which contribute to the effective
potential in the model of Sec.~4 at order $fg^4$ and $g^6$; the heavy
dots denote counterterm insertions.} and the $O(fg^4)$ and $O(g^6)$
counterterms.  The double-counting issue arises with the second graph
of Fig.~2, which has the same topology as the graph of Fig.~1c, but a
different (i.e., the standard) assigment of momenta.  However, since
the latter graph is $O(g^8)$ (after renormalization), the issue can be
ignored at this point.

      The sum of these graphs gives a contribution which, in terms of
an arbitrary renormalization scale $\tilde\sigma$, can be written in
the form
$$ \eqalign {V_2 (\phi) &=
    {1\over 2048\pi^4} \left[ f (m^2 + g^2 \phi^2)  + 32g^4\phi^2
      \right] (m^2 + g^2 \phi^2)
         \ln^2  \left( {m^2 + g^2 \phi^2
  \over m^2 + g^2 \tilde\sigma^2 } \right) \cr
     & \qquad  + {1\over 64\pi^2}(c_1 f + c_2 g^2)  (m^2 + g^2 \phi^2)^2
      \left[ \ln  \left( {m^2 + g^2 \phi^2
  \over m^2 + g^2 \tilde\sigma^2 } \right)  - {1\over 2} \right] \cr
      &\qquad  +  (c_3 f + c_4 g^2) m^2 (m^2 + g^2 \phi^2)
      \left[ \ln  \left( {m^2 + g^2 \phi^2
  \over m^2 + g^2 \tilde\sigma^2 } \right) -1 \right] \cr
   & \qquad    -{y_2 \over 4} \left( \phi^2 - \tilde\sigma^2 \right)^2
      + \left( {z_2\over 4!}  + {y_2\over 4} \right) \phi^4     }
     \eqno\eq  $$
Here the $c_k$ are numbers of order unity which depend on the precise
choice of the renormalization conditions fixing $m^2$, $g^2$, and the
$\chi$-field wave function renormalization, while $y_2$ and $z_2$ are
to be chosen so as to satisfy Eqs.~\mufix\ and \lambdafix.  If
$\tilde\sigma$ is again chosen to be $\sigma$, with the latter now
understood to be the minimum of the two-loop approximation to
$V_{\rm eff}$, this result combines with $V_{\rm 1-loop}$ to give
$$ \eqalign{  V_{\rm 2-loop}(\phi) &=
  {1\over 64\pi^2 }( 1 + c_1 f + c_2 g^2) (m^2 + g^2 \phi^2)^2
   \left[ \ln  \left( {m^2 + g^2 \phi^2
  \over m^2 + g^2 \sigma^2 } \right) - {1\over 2} \right] \cr
        &+ {1\over 2048\pi^4} \left[ f (m^2 + g^2 \phi^2)  + 32g^4\phi^2
      \right] (m^2 + g^2 \phi^2)
     \ln^2  \left( {m^2 + g^2 \phi^2
  \over m^2 + g^2  \sigma^2 } \right) \cr
      &+ (  c_3 f + c_4 g^2) m^2 (m^2 + g^2 \phi^2)
   \left[ \ln  \left( {m^2 + g^2 \phi^2
  \over m^2 + g^2 \sigma^2 } \right) - 1 \right] \cr
   & -{1\over 4}\left({\mu^2 \over \sigma^2} + y\right)
      \left( \phi^2 - \sigma^2\right)^2 }
    \eqno\eq $$
where $y= y_1 +y_2$.

       Among the contributions at $O(g^8)$ are those from the graphs
with a single $\phi$-loop.  These include a term proportional to
$\ln V_{\rm 1-loop}''(\phi)$ which is complex whenever $V_{\rm 1-loop}''$ is
negative.  While such complex effective potentials are not unfamiliar,
and their physical interpretation is
understood\rlap,\ref{E.~J.~Weinberg and A.~Wu, \pr{36}{2474}87. }
their implications for bubble nucleation calculations have remained
somewhat unclear.   We will see below how the potential problems
associated with these are avoided.

     Let us now turn to the bubble nucleation problem.  The first step
is to calculate $W(\phi)$.  Because the Lagrangian only contains even
powers of $\chi$, only one-particle irreducible graphs contribute.
One obtains
$$  \eqalign{ W(\phi) &= \int d^4x \,
   \left[ {1\over 2} \left(\partial_\mu\phi\right)^2
      + {1 \over 2} \mu^2 \phi^2 + {\lambda \over 4} \phi^4
      \right]
    + {1 \over 2}\int d^4x \,\langle x|
  \ln \left[ -\square + M^2(\phi)  \right] |x\rangle \cr
        & \qquad + {\rm counterterms} + \cdots }
       \eqno\eq $$
where the contributions from tree and one-loop graphs have been shown
explicitly.   The second term, arising from the loop graphs, is
sensitive to the value of $\phi(y)$ over a region of size $\sim
M^{-1}(\phi) \sim (g\sigma)^{-1} $.  Hence, for our purposes a slowly
varying $\phi$ will be one which varies slowly over this distance.
Expanding such a field about its value at a point $x$ leads to the
derivative expansion for $W$.

    The first term in the expansion, $\vh$, is obtained by
treating $\phi$ as a constant. To order $g^4$ this is the same as the
standard calculation of the effective potential, and gives the result
displayed in Eq.~\voneloop.  However, the two quantities differ at order
$g^6$, since the two-loop graph of Fig.~2b contributes to
$V_{\rm eff}$ but not to $\vh$.  Because this graph is divergent,
while the counterterm contributions to the two potentials are the
same, $\vh$, unlike $V_{\rm eff}$, cannot be finite.  This divergence is
acceptable because $\vh$ does not directly correspond to any
physical quantity.  Indeed, it combines with other
divergent quantities to give a finite result for $\Gamma$.

     The calculation of the next term in the derivative expansion is
equivalent to extracting the term of order $p^2$ from the sum of
graphs with two external $\phi$-lines carrying nonzero momentum $p$ and
any number of zero-momentum external $\phi$-lines.  Using dressed
propagators, the leading part of this sum can be represented by the
single graph of Fig.~4,
\FIG\four{A one-loop graph from which the leading contributions to
$\zh(\phi)$ can be computed.}
whose evaluation gives
$$ f(p^2) = f(0) + { g^4 \over 8 \pi^2 }
   \int_0^1 dx \ln \left[1 + { p^2x(1-x) \over M^2(\phi) } \right]
     \eqn\fofp $$
For $p^2 \simle M^2(\phi)$ this can be expanded
in powers of $p^2$:
$$ f(p^2) = f(0) + { g^4 \over 48 \pi^2 } \left[
{p^2 \over  M^2(\phi)} + {p^4 \over 5 M^4(\phi)} + \cdots \right]
     \eqno\eq $$
(The failure of this expansion when $p^2 \simge
M^2(\phi) $ reflects the failure of the derivative expansion when the
background $\phi$ field is not slowly varying.)  The first term,
$f(0)$, has been included in the calculation of $\vh$.  The
$O(p^2)$ term leads to
$$ \zh(\phi) = 1 + {g^2 \over 48\pi^2}
   \left[ {g^2\phi^2 \over m^2 + g^2\phi^2}
    + k  \right] + O(g^4)
      \eqno\eq $$
where $k$ is a constant of order unity which is chosen
to enforce the $\phi$ wave
function renormalization condition.  (Note that to this order
$\zh(\phi)= Z_\phi(\phi,0)$.)  The $O(p^4)$ term gives a contribution
to the derivative expansion of the form $ H(\phi) (\square \phi)^2 $.
In addition, there are other four-derivative terms, proportional to
$(\partial_\mu \phi)^2 \,\square \phi$ and $(\partial_\mu \phi)^2
(\partial_\nu \phi)^2$, which can obtained from graphs with three or
four external lines carrying nonzero momentum.

     The bounce solution is determined by the approximate action
of Eq.~\wzero\ with $\vh_{\rm 1-loop}(\phi)$ being just the
$V_{\rm 1-loop}(\phi)$ of Eq.~\voneloop.  Scaling arguments similar to
those of Sec.~II show that it has a spatial extent of order $(g^2
\sigma)^{-1}$ and thus varies slowly enough to allow a derivative
expansion.  By contrast, in a theory with couplings of ``normal''
magnitude ($\lambda \sim g^2$ and $\mu^2 \sim g^2 \sigma^2$) the
bounce would have a size $\sim (g\sigma)^{-1}$ and would not be slowly
varying.  Of course, in that case the loop corrections to the
effective potential would not change the vacuum structure, so the
standard calculation of $\Gamma$ could be used and there would be no
need for a derivative expansion.

    We need to know the magnitude of the various terms in
the derivative expansion of $W(\phi_\b)$.  Since $\phi_\b \sim
\sigma$, we see that $\vh(\phi_\b) \sim g^4 \sigma^4$, while our
estimate of the bounce size implies that $(\partial_\mu \phi_\b)^2 \sim
g^4 \sigma^4$. More generally, consider a term containing $j$
derivatives and $k$ explicit factors of $\phi_\b$, which might be
written schematically as
$$ w_{jk} = H_{jk}(\phi_\b) \partial^j \phi_\b^k
     \eqno\eq$$
$H_{jk}(\phi)$ must have dimensions of $(mass)^{4-j-k}$; since it is
derived from graphs with only internal $\chi$-lines, the mass
appearing here must be the effective $\chi$ mass, $M(\phi)$.  Further,
the structure of the graphs implies that there is a factor of $g$ for
each explicit factor of $\phi$, except in the tree-level contribution
to $\zh$.  Finally, each derivative brings in a factor of $g^2
\sigma$.  Thus, apart from the terms contributing to $\zh$ (i.e., the
case $j=k=2$),
$$  w_{jk}   \sim  \left({g^2 \sigma \over M(\phi) }\right)^j
      \left({g \phi \over M(\phi) }\right)^k
      M^4(\phi)
    \sim g^{4+j} \left({\phi\over \sigma}\right)^k \sigma^4
     \eqn\wjk$$
where the fact that $M^2(\phi) \ge m^2 \sim g^2\sigma^2$ has been used
and it has been assumed that $\phi \simle \sigma$.

      We can now estimate the various terms entering the formula
for $\Gamma$, beginning with the quantities $W_j$ in Eq.~\woneetc.
Because $\Gamma$ involves the differences between
these quantities evaluated at $\phi = \phi_\b$ and at $\phi
=\phi_{\fv}$, the integrals are effectively restricted to a region
of volume $ \sim (1/g^2\sigma)^4$.  Using the above estimates for
the integrands, we then find that the first two terms are
$$ W_0 = \int d^4x [ \vh_{g^4} + (\partial \phi)^2 ]  \sim g^{-4}
     \eqno \eq $$
and
$$ W_1 = \int d^4x [ \vh_{g^6} + \zh_{g^2}(\partial \phi)^2 ]
    \sim g^{-2}
     \eqno \eq $$
Here the subscripts indicate the order of the contributions to
$\vh$ and $\zh$, with $\vh_{g^4}$ being the same as $\vh_{\rm 1-loop}$,
and $\vh_{g^6}$ including both $g^6$ and $fg^4$ terms.
The next term, $W_2$, is of order unity and contains $\vh_{g^8}$
(arising from graphs with up to three loops), $\zh_{g^4}$ (arising
from one- and two-loop graphs) and the one-loop contributions to
the four-derivative terms.

    We also need expansions of the first two variational
derivatives of $W$.  In principle, the variations should be
performed before the expansion is carried out, although this
turns out not to matter for the first variation.  Thus, to
one-loop order
$$  W'(\phi;z) =  -\square \phi(z) + V'(\phi(z))
    + g^2\phi(z)\langle z| \left [ - \square + M^2(\phi)
\right]^{-1}
       |z\rangle  + {\rm counterterms}
     \eqno\eq $$
The leading contribution to the third, nonlocal, term is obtained
by taking $\phi$ to be constant.  This combines with
$V'$ to give ${\vh}'_{\rm 1-loop}$, so that the leading approximation to
$W'$ is indeed the variation of the leading approximation to $W$.
One can check that analogous statements hold at the next order,
thus verifying Eq.~\wprime.

     Up to one-loop order the second variational derivative is
$$ \eqalign {W''(\phi;z,z') &= \delta^{(4)}(z-z') [-\square + \vh''(\phi)]
      + g^2 \delta^{(4)}(z-z')
  \langle z| \left [ -\square + M^2(\phi) \right]^{-1}
       |z\rangle \cr
  &\qquad -2g^4 \phi(z) \phi(z')
   \langle z| \left[-\square + M^2(\phi)
  \right]^{-1}|z'\rangle ^2 + {\rm counterterms} }
     \eqn\chiwprimeprime $$
For $|z-z'| \simle M(\phi)$ the nonlocality of the third term
cannot be ignored and the derivative expansion fails, even in a
spatially constant background.  To see this,
let us suppose that $\phi(x)$ is constant, so that we can work in
momentum space with $p$ being the momentum conjugate to $z-z'$.  The last
term in Eq.~\chiwprimeprime\ is then given by the graph of Fig.~4.  If
$p$ were set equal to 0 in the evaluation of the loop, then the
one-loop terms shown here would give $W_0''$, which in momentum space
is simply $W_0''(p) = p^2 + \vh''_{\rm 1-loop}(\phi) = p^2 + {\cal
M}^2(\phi)$.  Subtracting this and recalling Eq.~\fofp, we see that
$\delta W''(p) \sim g^2 p^2 \ln [1+ p^2 / M^2 (\phi)]$.  If
$p^2/  M^2(\phi)$ is small, this can be treated as
as a small perturbation, but this is not the case when
$p^2/  M^2(\phi)$ is so large that the
logarithm overcomes the factor of $g^2$.  Correspondingly, in position
space $\langle z| [W_0'']^{-1} \delta W'' |z' \rangle$ is small
except in a region $|z  -z'| \ll M^{-1}(\phi)$.
Similar estimates clearly apply with a slowly varying background field.

     $W''$ enters the formula for the nucleation rate both through the
quantity $C_2$ defined by Eq.~\ctwo\ and through the determinant factor.
In both cases Eq.~\wprimeprimepert\ leads to a formal expansion in
which the higher order
terms are suppressed if $\delta W''$ is indeed a small perturbation;
i.e., if the contribution of the short-distance, high-momentum region
is suppressed.  Roughly speaking, $C_2$ can be viewed as
corresponding to a tree graph in which two factors of $\delta W'$ are
connected by a $\phi$ propagator.  The magnitude of the momentum
running through this propagator is set by the spatial extent of the
bounce solution, and is indeed small enough for $\delta W''$ to be
treated as a perturbation. (This can be checked by a detailed examination
of the contribution from the small $z-z'$ region.)  We may therefore
write
$$  C_2 = \int d^4z \, d^4z' \, W_1'(\phi_\b; z) W_1'(\phi_\b;z')
   \langle z| [W_0''(\phi_\b)]^{-1} |z' \rangle
      - (\phi_\b \rightarrow \phi_{\fv}) + \cdots
     \eqn\chictwo $$
The size of the bounce restricts the $z$ integration to a volume
of size $(1/g^2 \sigma)^4$, while the falloff of $[W_0'']^{-1}$
restricts $z-z'$ to a similar volume. The two factors of $W_1'$
are each of order $g^6 \sigma^3$, while $[W_0'']^{-1}$ gives a
factor of roughly $(z-z')^2$.  Combining these facts, one finds
that the terms shown explicitly in Eq.~\chictwo\ are of order unity.

      For the determinant factor, we recall from Eq.~\detexp\ that
$\det W''$ can be written as a product of $\det W_0''$ and the
exponential of a power series in $(W_0'')^{-1} \delta W''$.  The
former combines with the Jacobean to give $\sigma^4$ times a factor of
order unity, as in the standard calculation.  The latter corresponds
to a sum of one-loop graphs in which a number of insertions of $\delta
W''$ are connected by an equal number of $\phi$-propagators.  This sum
has an ultraviolet divergence which can be attributed to the graphs
with one and two insertions of $\delta W''$; even after the
cancellation of these divergences by counterterm contributions, we
should expect the high momentum region to be enhanced, and these terms
to be anomalously large.  This is in fact what happens.  Working in
momentum space, and ignoring the spatial variation of $\phi$ for the
sake of clarity, we may write the $n$th term in the expansion as
$$ \tr \left[ (W_0'')^{-1} \delta W'' \right]^n
    = \int d^4x\, d^4p \left\{ [W_0'']^{-1}(p) \delta W''(p) \right\}^n
        \eqno\eq $$
The momentum integration is quartically
divergent; after cancellation of infinities by counterterms a finite
term proportional to the fourth power of $M(\phi)$, the largest
relevant mass, remains.  There is a factor of $g^2$ for each $\delta
W''$.  Finally, because we must eventually take the difference between
the contributions of the bounce and of the pure false vacuum, the $x$
integration is effectively restricted to a volume $\sim (g^2
\sigma)^{-4}$.  The net result is a contribution of order $g^{2n-4}$,
implying that both the $n=1$ and the $n=2$ terms are of order unity or
larger, as indicated in Eq.~\detexp.

       The dominant, $O(g^{-2})$, contribution from the $n=1$ term is
obtained by treating $\phi$ as a constant.  This term then essentially
reproduces the two-loop graph of Fig.~2b, which contributes to
$V_{\rm eff}$ but not to $\vh$.  (The definition of $W_0''$ actually
requires a subtraction corresponding to the graph of Fig.~1c, but this
is simply the correction for double-counting noted previously.)  To
this order, then, the net effect is to simply replace $\vh$ by the
full effective potential.  Furthermore, since $\zh$ and $Z_\phi$ are
equal at $O(e^2)$, it might seem that the derivative expansion of $W(\phi)$
is simply being replaced by the derivative expansion of $S_{\rm eff}(\phi,
\chi=0)$.

     However, when we go to the next higher order it becomes clear
that this is not the case.  At $O(g^8)$, $V_{\rm eff}$ differs from $\vh$
by the contribution from two- and three- loop graphs with both $\chi$
and $\phi$ internal lines plus the contribution from graphs with a
single $\phi$-loop. The former set of graphs is reproduced by the
$n=2$ term in the expansion of the determinant and by the next to
leading contribution of the $n=1$ term.  The contribution of the
latter set is contained in the factors of $\det W_0''$; indeed, for a
constant background $\phi$ the logarithm of this determinant is just
the spatial integral of this part of the effective potential.
Matters are more complicated when this determinant is evaluated in the
background of the bounce.  Since the effective $\phi$ mass is a
factor of $g$ smaller than the effective $\chi$ mass, the derivative
expansion of $\det W_0''$ requires a more slowly varying background
field than is needed for the expansion of the terms in $W(\phi)$
arising from $\chi$-loops.  Precisely because the bounce is determined
by the field equations of $W_0$, its characteristic size is
incompatible with a derivative expansion of $\det W_0''(\phi_\b)$, and
so the $\phi$-loop contribution to $V_{\rm eff}$ cannot be simply
isolated.

      This resolves a puzzle associated with the
complexity of the perturbative effective potential.  As noted above,
the $\phi$-loop contribution to $V_{\rm eff}(\phi)$ is complex for values
of $\phi$ such that $V_{\rm 1-loop}''(\phi)$ is negative.  Since the
bounce solution $\phi_\b(x)$ lies in this range for some values of $x$,
a manifestly complex result for the nucleation rate would have been obtained
if the calculation had explicitly involved the full effective
potential\rlap.\footnote{4}{Although $\det W_0''(\phi_\fv)$ can be expanded
in a derivative expansion, no complexity problem arises because
$V_{\rm eff}(\phi_{\fv})$ is real.}

      Of course, the fact that this effective potential contribution
cannot be isolated does not rule out the possibility that a complexity
problem could arise with the full determinant, where it would be manifested
by the existence of a negative eigenvalue
of $W_0''(\phi_\b)$ in addition to the one which occurs for any bounce.
However, from the fact that the bounce corresponds to the solution of a
minimization problem (that of finding the path through configuration space
with the smallest WKB tunneling exponent), it is easy to
show\ref{S.~Coleman, The Uses of Instantons, in {\it The Whys of
Subnuclear Physics}, ed. A.~Zichichi (Plenum, 1979).}
that the bounce solution of minimum action can never have
additional negative modes.  Hence, $|{\det}' W_0''(\phi_\b)|^{1/2}$ is
real and the imaginary part of the effective potential has no effect on
the bubble nucleation calculation.

     To summarize these results, we may write
$$ \Gamma = \tilde A  \, e^{-(\bar B_0 + \bar B_1)}
     \eqn\gammafirst$$
where
$$ \bar B_0 = \int d^4x \left\{ \left[ V_{\rm eff}^{g^4}(\phi_\b)
     +  {1\over 2}(\partial_\mu\phi_\b)^2 \right]
     -(\phi_\b \rightarrow \phi_{\fv}) \right\} = O(g^{-4})
      \eqno\eq $$
$$ \bar B_1 = \int d^4x \left\{ \left[ V_{\rm eff}^{g^6}(\phi_\b)
     + {1\over 2} Z_\phi^{g^2}(\phi_\b) (\partial_\mu\phi_\b)^2 \right]
     -(\phi_\b \rightarrow \phi_{\fv}) \right\} = O(g^{-2})
      \eqno\eq $$
and the pre-exponential factor $\tilde A$,  now understood to include
the contribution from the $O(1)$ part of the exponent in
Eq.~\finalexp, is equal to $\sigma^4$ times a dimensionless factor of
order unity.

\vfill\eject

\chapter {Example 2: Scalar Electrodynamics}

     A second example is provided by scalar
electrodynamics\rlap.\refmark{\COLEMANWEINBERG} In terms of the real and
imaginary parts, $\phi_1$ and $\phi_2$ of the scalar field, the Lagrangian
takes the form $$ {\cal L} = -{1\over 4} F_{\mu\nu}^2
        + {1\over 2} \left(\partial_\mu\phi_1 + e A_\mu \phi_2 \right)^2
        + {1\over 2} \left(\partial_\mu\phi_2 - e A_\mu \phi_1 \right)^2
        - {1\over 2} \mu^2 \phi^2 - {\lambda\over 4!} \phi^4
    \eqno\eq $$
where $\phi^2 \equiv \phi_1^2 + \phi_2^2$.  In order that the one-loop
radiative corrections give rise to the desired vacuum structure with both
a symmetric minimum at $\phi=0$ and a
symmetry-breaking minimum at $\phi = \sigma$, we require that
$\lambda$ be $O(e^4)$ and that $\mu^2$ be positive and
$O(e^4\sigma^2)$.
As with the previous example, $\mu$ and $\lambda$ are to be fixed by
Eqs.~\mufix\ and \lambdafix, while the other renormalization
conditions will not be explicitly given.

       It is most convenient to calculate the effective potential in
Landau gauge.  In this gauge all graphs with a zero-momentum external
$\phi$-line entering a $\phi$-$\phi$-$A_\mu$ vertex vanish; this leads
to a considerable reduction in the number of graphs to be
calculated\rlap.\footnote{5}{The issue of gauge dependence is an
important one, with
some issues still to be settled.  It is known that the effective
potential is gauge-dependent beyond lowest order; this is acceptable
because it is not in general a directly measurable
quantity\rlap.\ref{For discussion of the gauge-dependence of the
effective potential and effective action, see R.~Jackiw, \pr9{1686}74;
R.~Fukuda and T.~Kugo,
\pr{13}{3469}76; N.~K.~Nielsen, \np{101}{173}75; I.~J.~R.~Aitchison
and C.~M.~Fraser, Ann. Phys. {\bf 156}, 1 (1984). }
Physically measurable quantities, on the other hand, should not depend
on the choice of gauge.  However, an
investigation\ref{N.~S.~Stathakis, Ph.D. Thesis, Columbia University, (1990)}
of the gauge-dependence of the bubble nucleation rate in scalar
electrodynamics found an apparent dependence on the gauge choice.}
A further
simplification follows from the observation that $V_{\rm eff}$ can only
depend on $\phi^2$, so that we may calculate it with $\phi_2=0$ and
$\phi_1=\phi$. To $O(e^4)$ the result is
$$ V_{\rm 1-loop}(\phi) =
   {1\over 2} \mu^2 \phi^2 + {\lambda\over 4!}  \phi^4
    + {3e^4\over 64 \pi^2} \phi^4
      \left[ \ln \left({\phi^2 \over \tilde\sigma^2} \right)
    - {25 \over 6}  \right]
          \eqno\eq $$
If, as before, we choose $\tilde \sigma = \sigma$, the minimum of
$V_{\rm 1-loop}$, we can eliminate $\lambda$ and, after adding a
constant,  write
$$  V_{\rm 1-loop}(\phi) = {3e^4\over 64 \pi^2} \phi^4
      \left[ \ln \left({\phi^2 \over \sigma^2} \right)
    - {1 \over 2}  \right]
     - {1\over 4} {\mu^2\over \sigma^2} \left(\phi^2 -\sigma^2\right)^2
   \eqno \eq $$
Requiring that $V''_{\rm 1-loop}(\sigma) > 0$ implies that $\mu^2 <
3e^4\sigma^2 / 16\pi^2$;
for this asymmetric minimum to be the true vacuum, the stronger
condition $\mu^2 < 3e^4\sigma^2 / 32\pi^2$ must hold.

     As in the previous example, the higher order contributions to the
effective potential are most easily obtained from Feyman graphs with
modified propagators.  In Landau gauge with $\phi_2=0$ these
are the dressed photon propagator with an effective photon mass
$$  M(\phi) = e \phi
   \eqno\eq $$
and the dressed scalar propagators with masses given by
$$ {\cal M}_1^2(\phi) = \left.{\partial^2 V_{\rm 1-loop} \over \partial
\phi_1^2}
        \right|_{\phi_1=\phi, \phi_2=0} = V_{\rm 1-loop}''(\phi)
    \eqno\eq $$
and
$$ {\cal M}_2^2(\phi) = \left.{\partial^2 V_{\rm 1-loop} \over \partial
\phi_2^2}
        \right|_{\phi_1=\phi, \phi_2=0} = \phi^{-1} V_{\rm 1-loop}'(\phi)
    \eqn\phitwomass $$
In gauges other than Landau gauge there would also be mixed
$\phi_2$-$A_\mu$ propagators.  As before, the inclusion of loops in
the dressed scalar propagators implies that caution must be used to
avoid double-counting.

    The $O(e^6)$ contributions to $V_{\rm eff}$ come from the
two-loop graphs of Fig.~5,
\FIG\five{The two-loop graphs which give $O(e^6)$ contributions to the
effective
potential of scalar electrodynamics.  Solid, dashed, and wiggly lines
refer to $\phi_1$, $\phi_2$, and photon propagators, respectively.}
one-loop counterterm graphs identical to those of Fig.~3, and
the $O(e^6)$ counterterms.  Combining these with the $O(e^4)$ terms,
with the renormalization point now chosen to be the minimum of the
two-loop approximation to the effective potential,
gives\ref{J.~S.~Kang, \pr{10}{3455}74.}
$$ \eqalign{ V_{\rm 2-loop}(\phi) &= {3e^4\over 64 \pi^2}( 1 + c
e^2) \phi^4
      \left[ \ln \left({\phi^2 \over \sigma^2} \right)
    - {1 \over 2}  \right]
      + {5 e^6 \over 512 \pi^4} \phi^4
         \ln^2 \left({\phi^2\over \sigma^2}\right) \cr
      &\qquad \qquad - {1\over 4} {\mu^2\over \sigma^2}
               \left(\phi^2 -\sigma^2\right)^2}
   \eqno\eq$$
where $c$, a number of order unity, depends on the precise specification of the
renormalization conditions.

     The contributions from purely scalar loops enter at $O(e^8)$.
These make the effective potential complex, since ${\cal M}_1^2$ is
negative in the region where $V_{\rm 1-loop}'' < 0$, while ${\cal M}_2^2 <
0$ when $|\phi| < \sigma$.

      As in the previous example, the function
$Z_\phi(\phi)$ will also enter the calculation of $\Gamma$.  At
$O(e^2)$ this receives contributions from the graphs of Fig.~6,
\FIG\six{The one-loop graphs which give $O(e^2)$ contributions to
$Z_\phi(\phi)$ in scalar electrodynamics.}
and is given by
$$   Z_\phi(\phi) = {3 e^2 \over 16\pi^2} \left[ \ln \left({\phi^2\over
\sigma^2}\right)
         + k \right]
      \eqno\eq $$
where $k$, of order unity, enforces the wave function renormalization
condition.

     Let us now turn to the calculation of the bubble nucleation rate.
Before going into the details, we can use the example of the previous
section to anticipate a number of the results.  The bounce is
presumably determined by field equations involving $V_{\rm 1-loop}$.  It
is then easy to see that it must have a constant phase, and so by a
global gauge rotation can be made entirely real.  Furthermore, the
spatial extent of the bounce must be of order $(e^2\sigma)^{-1}$, thus
making it slowly-varying enough to justify a derivative expansion of
the photon-loop terms, but not of those due to graphs with purely
scalar loops.  This will make it possible for the $O(e^6)$ part of the
effective potential to enter the calculation of $\Gamma$ in a
straightforward fashion, while still avoiding the troublesome $O(e^8)$
terms.

     There are at least two possible approaches to the calculation.  Since
the symmetry breaking arises from radiative corrections involving photon
loops, the most straighforward procedure would be to integrate over
the gauge field to obtain an action $W(\phi_1, \phi_2)$.  Since
$A_\mu$ enters the Lagrangian at most quadratically, this integration
can be done exactly, yielding
$$  \eqalign{ W(\phi_1, \phi_2) &= \int d^4x \,
   \left[ {1\over 2} \left(\partial_\mu\phi\right)^2
      + {1 \over 2} \mu^2 \phi^2 + {\lambda \over 4} \phi^4
      \right] \cr
    &\qquad    +  {1 \over 2}\tr \int d^4x \,\langle x|
  \ln \left [-\delta_{\mu\nu} \square  + \partial_\mu\partial_\nu
          +\delta_{\mu\nu}M^2(\phi)  \right] |x\rangle \cr
       & \qquad -{1\over 2} \int d^4x\, d^4x'\, j^\mu(x) j^\nu(x')
     \langle x| \left [-\delta_{\mu\nu} \square  + \partial_\mu\partial_\nu
          +\delta_{\mu\nu}M^2(\phi)  \right]^{-1} |x'\rangle \cr
        & \qquad + {\rm counterterms} }
       \eqno\eq $$
where $j_\mu = e( \phi_2 \partial_\mu\phi_1 - \phi_1 \partial_\mu \phi_2)$
and the factors of $\delta_{\mu\nu}$ arise from the Euclidean metric.
The second term on the right hand side corresponds to graphs with a
single photon loop, and is analogous to the one $\chi$-loop term in the model
of Sec.~IV.  The last term corresponds to a graph with a photon line
connecting two currents.  While it vanishes for both the constant false
vacuum solution and the bounce,  it does not vanish in
general and, indeed, it is clearly needed to bring the $O(e^6)$ part
of the effective potential into the exponent of $\Gamma$.  However,
because it cannot be expanded as a sum of local terms, the analysis of
this term introduces complications not encountered in the previous
example.

    These complications are avoided by an alternative
approach.  Since the bounce can be chosen to be entirely real, it
should be possible to integrate out $\phi_2$, as well as $A_\mu$, from
the very beginning to obtain an effective action
$$  W(\phi) = \int d^4x \,
   \left[ {1\over 2} \left(\partial_\mu\phi\right)^2
      + {1 \over 2} \mu^2 \phi^2 + {\lambda \over 4} \phi^4
      \right]   +  {1\over 2} \tr \ln G  + {\rm counterterms} + \cdots
     \eqno\eq$$
Here the dots represent multiloop contributions while
$G$ is the matrix
$$ G = \left.\left(\matrix{ {\delta^2 S \over \delta A_\mu \delta A_\nu} &
     {\delta^2 S \over \delta A_\mu \delta \phi_2} \cr
     {\delta^2 S \over \delta \phi_2 \delta A_\nu} &
     {\delta^2 S \over \delta \phi_2^2 } \cr} \right)\right|_{\phi_1=\phi,
  \phi_2=A_\mu=0}
    \equiv \left(\matrix { G_{\mu\nu}(\phi) & G_{\mu 2}(\phi) \cr
    G_{2\nu}(\phi) & G_{2 2}(\phi) \cr} \right)
     \eqno\eq $$
In Landau gauge with a constant background $\phi_1(x)$ the
off-diagonal entries of this matrix vanish.  For a nonconstant but
slowly-varying background one can write
$$  G = \left(\matrix { G_{\mu\nu}(\phi) & 0 \cr
    0  & G_{2 2}(\phi) \cr }\right)
    \left[ I + \left(\matrix {0& G_{\nu\mu}^{-1}(\phi)  G_{\mu 2}(\phi) \cr
    G_{22}^{-1}(\phi)  G_{2 \nu}(\phi) &0\cr }\right)
    \right]
     \eqno\eq $$
and then expand $W$ as
$$ \eqalign { W &= \int d^4x \,
   \left[ {1\over 2} \left(\partial_\mu\phi\right)^2
      + {1 \over 2} \mu^2 \phi^2 + {\lambda \over 4} \phi^4
      \right] \cr  &\qquad
   + {1\over 2} \tr \ln G_{\mu\nu} + \ln\det G_{22}
    + {1\over 2} \tr G_{\nu\mu}^{-1} G_{\mu 2} G_{22}^{-1} G_{2 \nu }
        + {\rm counterterms} +\cdots }
      \eqno\eq$$
The first (tree-level) term and the second term, corresponding to
one-photon-loop effects, dominate.  Aside from a minor point noted below,
these can be treated just like the analogous terms for the previous
model.  Let us write these as $W_0 + \delta_1W$, with $W_0$, given by
Eq.~\wzero, again being the approximate action which determines the bounce
solution $\phi_\b(x)$.  The spatial extent of the bounce is such that
$\delta_1W(\phi_\b)$ can be expanded in a derivative expansion.  This
expansion does not give any further contribution to $\vh$, but does
give the contribution to $\zh$ corresponding to the graph of Fig.~6a.

    Let us consider next the
fourth term, which may be denoted $\delta_2W$.  This corresponds to
one-loop graphs with both a photon propagator and a $\phi_2$-propagator.
Because $G_{22}$ is obtained from the fundamental action $S$, the latter
propagator will have a mass $\hat {\cal M}_2(\phi)$ given by the tree-level
potential  rather than by $V_{\rm 1-loop}$, as is
the mass ${\cal M}_2(\phi)$ of Eq.~\phitwomass.  The replacement
$\hat{\cal M}_2 \rightarrow {\cal M}_2 $ is obtained by summing over
multiloop graphs with arbitrary numbers of photon loops attached to the
$\phi_2$-line, as in Fig.~7;
\FIG\seven{Multiloop graphs whose contributions must be added to
$\delta_2W(\phi)$.}
let us denote the result of this replacement
by $\hat\delta_2W(\phi)$.  (Since both $\hat {\cal M}_2$ and ${\cal M}_2$
are a factor $e$ smaller than the effective photon mass,
$\hat\delta_2W(\phi)$ and $\delta_2W(\phi)$ actually differ only in
subdominant terms.)  About the bounce solution $\hat\delta_2W(\phi)$ can
also be expanded in a derivative expansion, with the leading term being
the contribution of graph 7b to $\zh(\phi)$.

       The third term, $\ln \det G_{22}(\phi)$, is the contribution from
one-loop graphs with only internal $\phi_2$ propagators.  As with
$\delta_2W$, these propagators have the mass $\hat {\cal M}_2$.  Summing
over graphs with photon loops attached to the $\phi_2$-line again
replaces this mass by ${\cal M}_2$ and replaces $G_{22}(\phi)$ by
$$ \bar G_{22}(\phi) = -\square + {\cal M}^2_2(\phi)
     \eqno \eq$$
Two points should be noted here.  First, the effects of the change of mass
are not subdominant, as they were in the previous case.  Second, this
term cannot be expanded in a derivative expansion about the bounce.

    Doing the path integral about the bounce and the pure false
vacuum and then proceeding as in the previous example leads to
$$  \Gamma = e^{-B_0} \, e^{-B_1} \, e^{-B_2}
    \left| {{\det}' W_0''(\phi_\b) \over {\det} W_0''(\phi_{\fv}) }
      \right|^{-1/2} \,
    \left( {{\det}{\bar G}_{22}''(\phi_\b) \over {\det}
{\bar G}_{22}''(\phi_{\fv}) }
      \right)^{-1/2} \,
        {B^2_0 \over 4\pi^2} \, (1 + \cdots)
      \eqno\eq $$
where
$$  B_0 = W_0(\phi_\b) - W_0(\phi_{\fv})
       \eqno\eq $$
$$ \eqalign { B_1 &= \delta_1W(\phi_\b) + \delta_2W(\phi_\b) +
      \tr [W_0''(\phi_\b)]^{-1} \delta_1 W''(\phi_\b) \cr
  & \qquad   +  \tr [W_0''(\phi_\b)]^{-1} \delta_2 W''(\phi_\b)
   - (\phi_\b \rightarrow \phi_{\fv} )  }
       \eqno\eq $$
and $B_2$, containing higher order terms, is of order unity.  The
leading term in the exponent, $B_0$, contains the $e^4$ part of the
effective potential and the tree level kinetic term.  The four terms
in $B_1$ reproduce, in the order in which they appear, graphs 7a and
7b of the $O(e^2)$ part of $Z_\phi$ and graphs 5a and 5b of the
$O(e^6)$ part of $V_{\rm eff}$.  Although some of the $O(e^8)$ terms in
$V_{\rm eff}$ are contained in $B_2$, the potentially complex ones,
arising from scalar loops, are contained entirely within the explicit
determinant factors.  Because these determinants cannot be expanded in
derivative expansions, the imaginary part of the effective potential
does not explicitly enter the calculation.  Just as before, it is
necessary to look for negative eigenvalues in the spectra of the
operators in these determinants.  By viewing the bounce as the optimum
tunneling path in the space of field configurations involving both
$\phi_1$ and $\phi_2$, it is clear that there is still just a single
negative mode, involving only $\phi_1$.  However, in addition to the
four translational modes there is a fifth zero-frequency mode,
corresponding to a phase rotation of the bounce.   This must be
treated by introducing a collective coordinate, replacing $\det
\bar G_{22}(\phi_\b)$ by ${\det}'\bar G_{22}(\phi_\b)$, and multiplying by
a Jacobean factor
$$  J_{\rm phase} = \left[ 2 \pi \int d^4x \, \phi_\b^2(x) \right]^{1/2}
     \sim {1 \over e^4\sigma }
     \eqno\eq $$

     There remains a minor point, which was noted above.
Because the photon mass $M(\phi)$ is proportional to $\phi$, there can
be infrared problems if $\phi \ll \sigma$, as
is the case far from the center of the bounce, where $\phi_\b(x)$
exponentially approaches the false vacuum $\phi_{\fv}=0$.  These can be
seen most clearly by recalling the estimate \wjk\ for the effective
action contribution containing $j$ derivatives and $k$
factors of $\phi$.  With $M(\phi) = e\phi$, the previous estimate is
replaced by
$$ w_{jk} \sim g^{4+j} \left(\sigma\over \phi\right)^{j-4} \sigma^4
    \eqno\eq$$
For small $\phi$, the derivative expansion cannot be carried
out beyond the four-derivative terms.  However, if we extract from $W$ the
potential terms and the two-derivative terms, the remainder is
$O(g^4)$ and gives an $O(1)$ contribution to the exponent $B$,
which can eventually be absorbed in a redefinition of the prefactor.

       To summarize, the bubble nucleation rate is again given by an
expression of the form of Eq.~\gammafirst, with
$$ \bar B_0 = \int d^4x \left\{ \left[ V_{\rm eff}^{e^4}(\phi_\b)
     +  {1\over 2}(\partial_\mu\phi_\b)^2 \right]
     -(\phi_\b \rightarrow \phi_{\fv}) \right\} = O(e^{-4})
      \eqno\eq $$
and
$$ \bar B_1 = \int d^4x \left\{ \left[ V_{\rm eff}^{e^6}(\phi_\b)
     + {1\over 2} Z_\phi^{e^2}(\phi_\b) (\partial_\mu\phi_\b)^2 \right]
     -(\phi_\b \rightarrow \phi_{\fv}) \right\} = O(e^{-2})
      \eqno\eq $$
and the pre-exponential factor $\tilde A$ now being of order
$\sigma^4/e^4$.

\vfill\eject

\chapter{Concluding Remarks}

    In this paper I have shown how the decay rate of a metastable
vacuum can be calculated in a theory whose vacuum structure is
determined by radiative corrections.  As in the standard case, the
result may be written as a dimensionful prefactor times the
exponential of an action involving a bounce solution.  To leading
approximation this exponent is just the tree-level action supplemented
by the dominant one-loop contribution to the effective potential; in
scalar electrodynamics it is $O(1/e^4)$.  The first
correction to the exponent arises from the next-to-leading contributions to
the effective potential and the leading correction to the tree-level
kinetic part of the effective action.  Although smaller than the leading
terms, these give an addition to the exponent which is larger than order
unity (e.g., it is $O(1/e^2)$ in scalar electrodynamics) and is thus
more important than the prefactor.  It does not appear that this
correction need have any particular sign, but rather that it might
increase the nucleation rate in some theories and reduce the rate in
others.  Further, the separation between the leading and
next-to-leading terms in the bounce action is dependent on the precise
specification of the renormalization conditions.  In fact, the
existence of a correction term of this magnitude follows simply from
the requirement that the physical nucleation rate be
independent of the renormalization scheme and of the particular
definition of the coupling constants of the theory.

      All further corrections  may be
absorbed into the prefactor.  Although some of these (those
corresponding to graphs with internal lines of the heavy
``$\chi$-particles'') can be identified with particular terms in the
effective potential and the other functions entering the effective
action, this is not true of all the higher corrections.  Specifically,
the $\phi$-loop graphs which give rise to complex terms in
the effective potential cannot, when calculated in the background of
the bounce, be expanded in a derivative expansion.  Consequently, the
imaginary part of the effective potential does not explicitly enter
the bubble nucleation calculation, and the problems of interpretation
which it would entail are avoided.  In particular, the expression
obtained for the nucleation rate can be shown to be real.

      In the examples considered in this paper, it was obvious that
the standard formalism for vacuum decay had to be modified, since the
tree-level potential had only a single vacuum and so could not
possibly lead to a bounce solution.
However, one could easily construct examples in which the tree-level
potential had several inequivalent minima, but where the one-loop
corrections to the effective potential were comparable to the
tree-level terms and changed the vacuum structure.  In such cases a
bounce solution to the tree-level action would exist and could be used
to calculate the nucleation rate, but the result would not be
correct\rlap.\footnote{6}{A similar situation could arise in the study
of solitons.  The calculation of the quantum
corrections to the soliton energy has many similarities to the decay rate
calculation considered in this paper.  Here too, the standard
calculation should break down if the one-loop corrections to the
effective potential are large enough.}
Working with the standard formalism, would one see any indication that
the calculation was unreliable?  The answer is yes. To see this, consider a
model
similar to that of Sec.~4, but with an additional $\phi^3$ interaction
of magnitude such that $V(\phi, \chi=0)$ has two minima.
Further, assume that the $\phi$ self-interactions can be written in
the form
$$ V(\phi, \chi=0) = \lambda \sigma^4 U(\phi/\sigma)
        \eqno\eq $$
with $U$ involving no small couplings and the relation between
$\lambda$ and the couplings involving $\chi$ being the same as in
Sec.~4.  Applying the standard formalism in this case, one will
obtain a bounce solution with nontrivial $\phi(x)$ but with $\chi=0$
everywhere and will be led to an
expression for the nucleation rate which differs from Eq.~\oldanswer\
only by a multiplicative factor
$$  K_\chi  =   \left[{ {\det}[- \square + M^2(\phi_\b(x))] \over
     {\det}[- \square +M^2(\phi_{\fv})]} \right]^{-1/2}
 \eqno\eq$$
arising from the functional integration over $\chi$.

     The magnitudes of the various terms entering the expression for
$\Gamma$ can  then
be estimated by the methods of Sec.~2.  In particular,
since the spatial extent of the bounce solution is of order
$1/(\sqrt{\lambda}\, \sigma)$, it is useful to define the dimensionless
variable $s= \sqrt{\lambda}\, \sigma x$, in terms of which the bounce
has spatial extent of order unity.  The bounce action is then seen
to be of order $1/\lambda$, while the
determinants arising from the functional integration over $\phi$
combine with the Jacobean to give a factor of order unity.   However,
the remaining factor, $K_\chi$, now becomes
$$  K_\chi  =   \left[{ {\det}[- \square_s +
      M^2(\phi_\b(s))/\lambda\sigma^2 ] \over
     {\det}[- \square_s +M^2(\phi_{\fv})/\lambda\sigma^2]} \right]^{-1/2}
 \eqno\eq$$
Because the $\chi$-$\phi$ couplings have been assumed to be large
compared to the $\phi$
self-interactions, $M^2(\phi)/\lambda\sigma^2$ is large (of order
$1/g^2$).   As a result, $K_\chi$ is not simply a factor
of order unity which can be included in the prefactor. Instead, it
gives potentially large corrections whose evaluation\ref{For a recent
discussion of the evaluation of such determinant factors, see
S.~Dodelson and B.-A.~Gradwohl, FERMILAB-Pub-92/281-A (1992).}
is nontrivial; in other words, the standard algorithm has broken down.

     One final note.  This work has been concerned solely with the
problem of bubble nucleation by quantum mechanical tunneling at zero
temperature.  Issues similar to those encountered here also arise in
connection with the problem of finite temperature bubble nucleation,
even in theories where radiative corrections have little effect on the
zero-temperature vacuum structure.  These will be considered elsewhere.

\ack

    I am grateful to Nicholas Stathakis for many helpful
conversations.  I would also like to acknowledge the hospitality of
the Institute for Theoretical Physics, Santa Barbara, where part of
this work was done.

\vfill\eject

\refout
\vfill\eject
\figout
\end